\documentclass[11pt]{article}% %
\usepackage[width=15cm,height=23cm]{geometry} 
\usepackage{amsmath}
\usepackage{amssymb}
\usepackage{subfigure}
\usepackage[dvips]{graphicx}

\long\def\symbolfootnote[#1]#2{\begingroup%
\def\thefootnote{\fnsymbol{footnote}}\footnote[#1]{#2}\endgroup}

\newcommand{\comment}[1]{}

\begin{document}
\def \NLSE {nonlinear Schr\"{o}dinger equation}
\def \beq {\begin{equation}}
\def \eeq {\end{equation}}
\def \bea {\begin{eqnarray}}
\def \eea {\end{eqnarray}}
\def\opex{ Opt.\ Express }
\def\ao{ Appl.\  Opt.\ }
\def\ap{ Appl.\  Phys.\ }
\def\apa{ Appl.\  Phys.\ A }
\def\apb{ Appl.\  Phys.\ B }
\def\apl{ Appl.\ Phys.\ Lett.\ }
\def\apj{ Astrophys.\ J.\ }
\def\bell{ Bell Syst.\ Tech.\ J.\ }
\def\jqe{ IEEE J.\ Quantum Electron.\ }
\def\jlt{ J.\ Lightwave\ Technol.\ }
\def\ptl{ IEEE Photon.\ Tech.\ Lett.\ }
\def\assp{ IEEE Trans.\ Acoust.\ Speech Signal Process.\ }
\def\aprop{ IEEE Trans.\ Antennas Propag.\ }
\def\mtt{ IEEE Trans.\ Microwave Theory Tech.\ }
\def\iovs{ Invest.\ Ophthalmol.\ Vis.\ Sci.\ }
\def\jcp{ J.\ Chem.\ Phys.\ }
\def\jmo{ J.\ Mod.\ Opt.\ }
\def\josa{ J.\ Opt.\ Soc.\ Am.\ }
\def\josaa{ J.\ Opt.\ Soc.\ Am.\ A }
\def\josab{ J.\ Opt.\ Soc.\ Am.\ B }
\def\jpp{ J.\ Phys.\ (Paris) }
\def\nat{ Nature (London) }
\def\oc{ Opt.\ Commun.\ }
\def\ol{ Opt.\ Lett.\ }
\def\pl{ Phys.\ Lett.\ }
\def\pra{ Phys.\ Rev.\ A }
\def\prb{ Phys.\ Rev.\ B }
\def\prc{ Phys.\ Rev.\ C }
\def\prd{ Phys.\ Rev.\ D }
\def\pre{ Phys.\ Rev.\ E }
\def\prl{ Phys.\ Rev.\ Lett.\ }
\def\rmp{ Rev.\ Mod.\ Phys.\ }
\def\pspie{ Proc.\ SPIE\ }
\def\sjqe{ Sov.\ J.\ Quantum Electron.\ }
\def\vr{ Vision Res.\ }
\def\cleo{ {\it Conference on Lasers and Electro-Optics }}
\def\assl{ {\it Advanced Solid State Lasers }}
\def\tops{ Trends in Optics and Photonics }

\vskip 0.4 truecm
%\Large

\title{%
Beam Quality Factor of Single-Mode Gain-Guided Fiber Lasers}

\author{Krishna~Mohan~Gundu, Parisa~Gandomkar~Yarandi, and Arash~Mafi\\
\\
\small\em Department of Electrical Engineering and Computer Science\\
\small\em University of Wisconsin-Milwaukee\\
\small\em Milwaukee, WI, 53211}
\date{\today}
\maketitle

\begin{abstract}
The beam quality factor $M^2$ for the fundamental LP01 mode of a step-index fiber
is calculated in the presence of gain, in a closed form, as a function of the
complex generalized V-number. It is shown that the $M^2$ value of a
single-mode gain-guided fiber laser can be arbitrary large.
The results are important for the interpretation of the
beam quality measurements in recent experiments on single-mode gain-guided fiber lasers.
\end{abstract}

%\begin{center}{\Large\bf
%Beam Quality Factor of Single-Mode Gain-Guided Fiber Lasers
%}
%\end{center}
%\par
%\normalsize
%\large
%\vskip 1.1 truecm

%\large
%\rm
%\vskip 0.7 truecm
%\centerline{
%Krishna~Mohan~Gundu, Parisa~Gandomkar~Yarandi, and Arash~Mafi$^{}$
%}

%\normalsize
%\small
%\medskip
%\begin{center}
%Department of Electrical Engineering and Computer Science\\
%University of Wisconsin-Milwaukee\\
%Milwaukee, WI, 53211
%\end{center}

%
%\normalsize

%\vskip 1.3  truecm

%\centerline{Abstract}
%\begin{quotation}%
%
%\rm
% {\Large Abstract:}
%The beam quality factor $M^2$ for the fundamental LP01 mode of a step-index fiber
%is calculated in the presence of gain, in a closed form, as a function of the
%complex generalized V-number. It is shown that the $M^2$ value of a
%single-mode gain-guided fiber laser can be arbitrary large.
%The results are important for the interpretation of the
%beam quality measurements in recent experiments on single-mode gain-guided fiber lasers.
%\end{quotation}

%%%%%%%%%%%%%%%%%%%%%%%%%%%%%%%%%%%%%%%%%%%%%%%%%%%%
%%%%%%%%%%%%%%%%%%%%%%%%%%%%%%%%%%%%%%%%%%%%%%%%%%%%
%%%%%%%%%%%%%%%%%%%%%%%%%%%%%%%%%%%%%%%%%%%%%%%%%%%%
%%%%%%%%%%%%%%%%%%%%%%%%%%%%%%%%%%%%%%%%%%%%%%%%%%%%
%%%%%%%%%%%%%%%%%%%%%%%%%%%%%%%%%%%%%%%%%%%%%%%%%%%%
%\newpage
\section{Introduction}
\noindent There
has been a growing interest in optical fiber lasers that operate based on the
gain-guiding index-antiguiding (GG+IAG) principle. Unlike a conventional index-guiding (IG) fiber,
the core of an IAG fiber has a lower refractive index than the surrounding cladding and
cannot support IG modes. Siegman~\cite{Siegman1} has 
shown that in the presence of sufficient gain, an IAG fiber can support confined propagating GG modes, which
are normalizable in the transverse direction. More interestingly, even for arbitrarily large
core diameters, GG+IAG fibers can operate in a robust single transverse mode. It is therefore desirable to take 
advantage of the GG+IAG principle and scale up the core size while maintaining the 
single-mode characteristic. 
The large core size is attractive because it can help mitigate the unwanted nonlinear optical effects, raise the 
optical damage threshold, and increase the amplification per unit length of the fiber.

Several experiments have demonstrated GG+IAG in various fiber laser configurations~\cite{Siegman5,McComb1,Chen1,Chen2}. The reported values of the 
beam quality factor $M^2$ in these experiments are notably larger than unity, even in fibers that are designed 
to operate as single-mode. This should not be surprising 
considering the substantial departure of the LP01 profile of a GG+IAG fiber from a 
Gaussian-like beam~\cite{Siegman2}, as similarly reported  
in other unconventional optical fibers~\cite{Mafi1}.
However, we show that the values of $M^2$ for the single-mode GG+IAG fibers are substantially larger those 
measured in the experiments. Therefore, the not-so-puzzling larger than unity $M^2$ measurements
in these fibers turn out to be quite lower than expected from a pure GG+IAG structure.

We present a closed form expression for the $M^2$ parameter of the LP01 mode 
in the presence of gain as a function of the complex generalized fiber V-number. 
The $M^2$ parameter has been calculated analytically for the LPmn modes in passive 
fibers~\cite{Yoda1}. However, we are not aware of a derivation in the presence of gain, which 
requires integrals of complex variables and is mathematically more
involved. In this paper, we only present the final results and details of the 
analytical derivation will be reported elsewhere.
\section{Gain-Guided Optical Fibers}
In order to study step-index optical fibers in the presence of gain (GG+IG or GG+IAG), it is 
convenient to use a generalized complex V parameter squared~\cite{Siegman1} defined as
\begin{equation}
\label{eq:Vtilde_definition}
{\tilde V}^2=\Delta N + i G.
\end{equation}
The index and gain parameters $\Delta N$ and $G$ 
are given by
\begin{align}
&\ \Delta N=\left(2\pi a/\lambda\right)^2 2n_0\Delta n,\\
&\ G=\left(2\pi a/\lambda\right)^2 \left(n_0\lambda/2\pi\right)g,
\end{align}
where $n_0$ is the refractive index of the cladding. $n_0+\Delta n$ is the refractive index of 
the core, $a$ is the core radius, $g$ is core power-gain coefficient, and $\lambda$ is the vacuum 
wavelength. For a proper choice of $\Delta N$ and $G$, the core can support an LP01 guided mode 
in the form of 
\begin{equation}
\label{eq:E_definition}
E(x,y,z_0)=
\begin{cases}
    {\tilde N} J_0(ur/a)/J_0(u),\qquad\ \ \ r\le a\\
    {\tilde N} K_0(wr/a)/K_0(w),\qquad r\ge a 
\end{cases}
\end{equation}
The parameters $u$ and $w$ are complex and satisfy the
following two equation, which can be used to determine 
these parameters, given ${\tilde V}$.  
\begin{align}
uJ_1(u)/J_0(u)&=wJ_1(w)/J_0(w),\\
u^2+w^2&={\tilde V}^2.
\end{align}
${\tilde N}$ is an overall constant to be determined from the 
normalization condition (\ref{eq:normalization}) assumed throughout this paper,
\begin{equation}
\label{eq:normalization}
\iint dS\left|E(x,y,z)\right|^2=1,
\end{equation}
where $\iint dS\overset{\underset{\mathrm{def}}{}}{=}\iint dxdy$.
In general, the value of
${\tilde V}$ determines the total number of confined guided modes in a fiber in the 
presence of gain, which can be zero or higher. The single-mode operating regions of the 
GG+IG and GG+IAG structures in the $\Delta N$-$G$ Cartesian space are presented in detail by 
Siegman~\cite{Siegman1,Siegman2} and will not be repeated here.
\section{Beam Quality Factor $M^2$}
The beam quality factor $M^2$ is commonly
used in experiments to determine whether a fiber laser is single-mode 
or not. For single-mode fiber lasers, the value of $M^2$ is near unity, while
$M^2>1$ is indicative of beam contamination with higher order modes.
In order to calculate the $M^2$, we adopt the variance method, which 
is mathematically rigorous and
closely resembles the common experimental procedures for the $M^2$ measurement \cite{Johnston1}.

\begin{figure}[htp]
  \begin{center}
    \subfigure[]{\label{fig:M21}\includegraphics[width=2.5in]{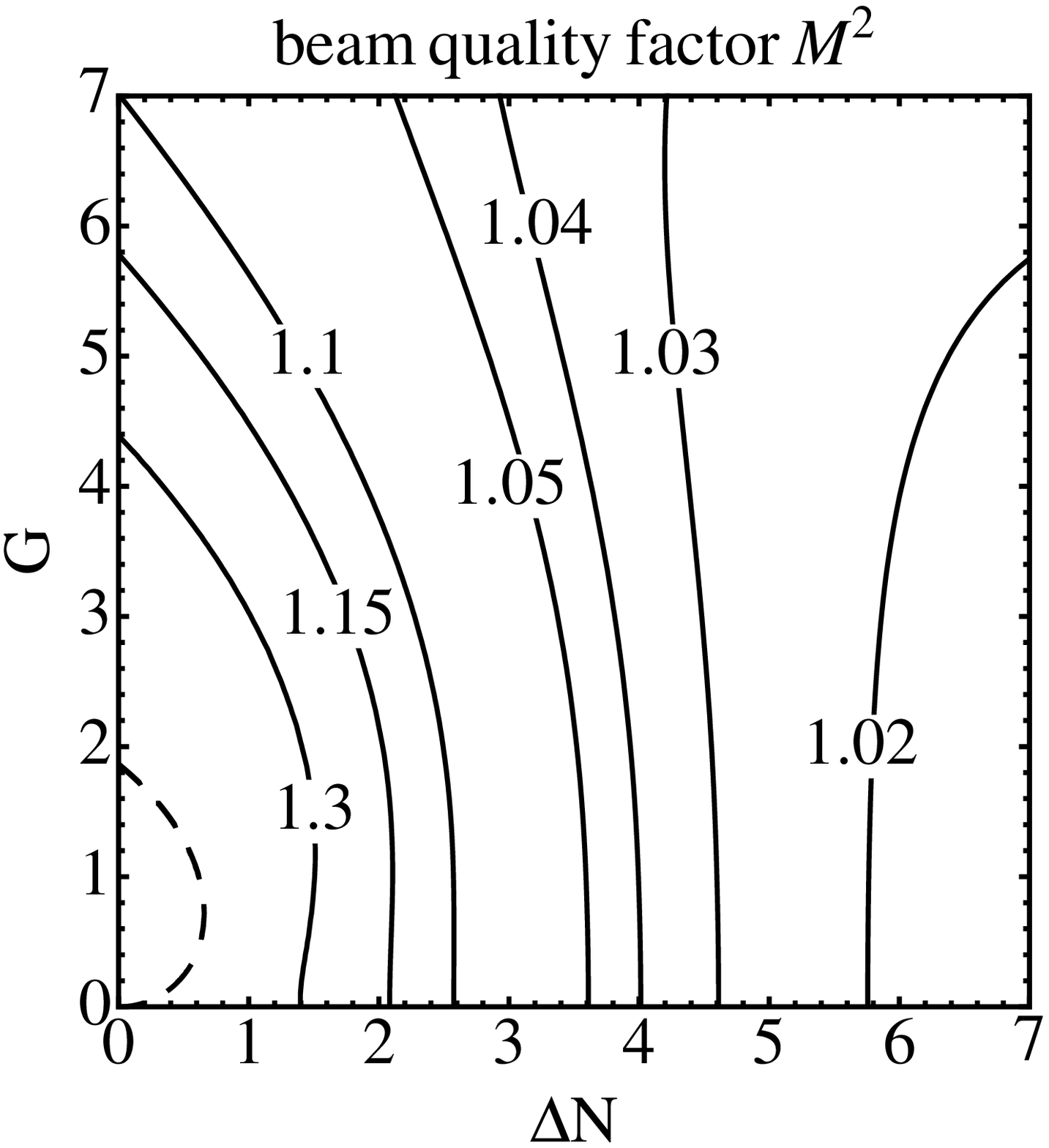}}
    \subfigure[]{\label{fig:M22}\includegraphics[width=2.6in]{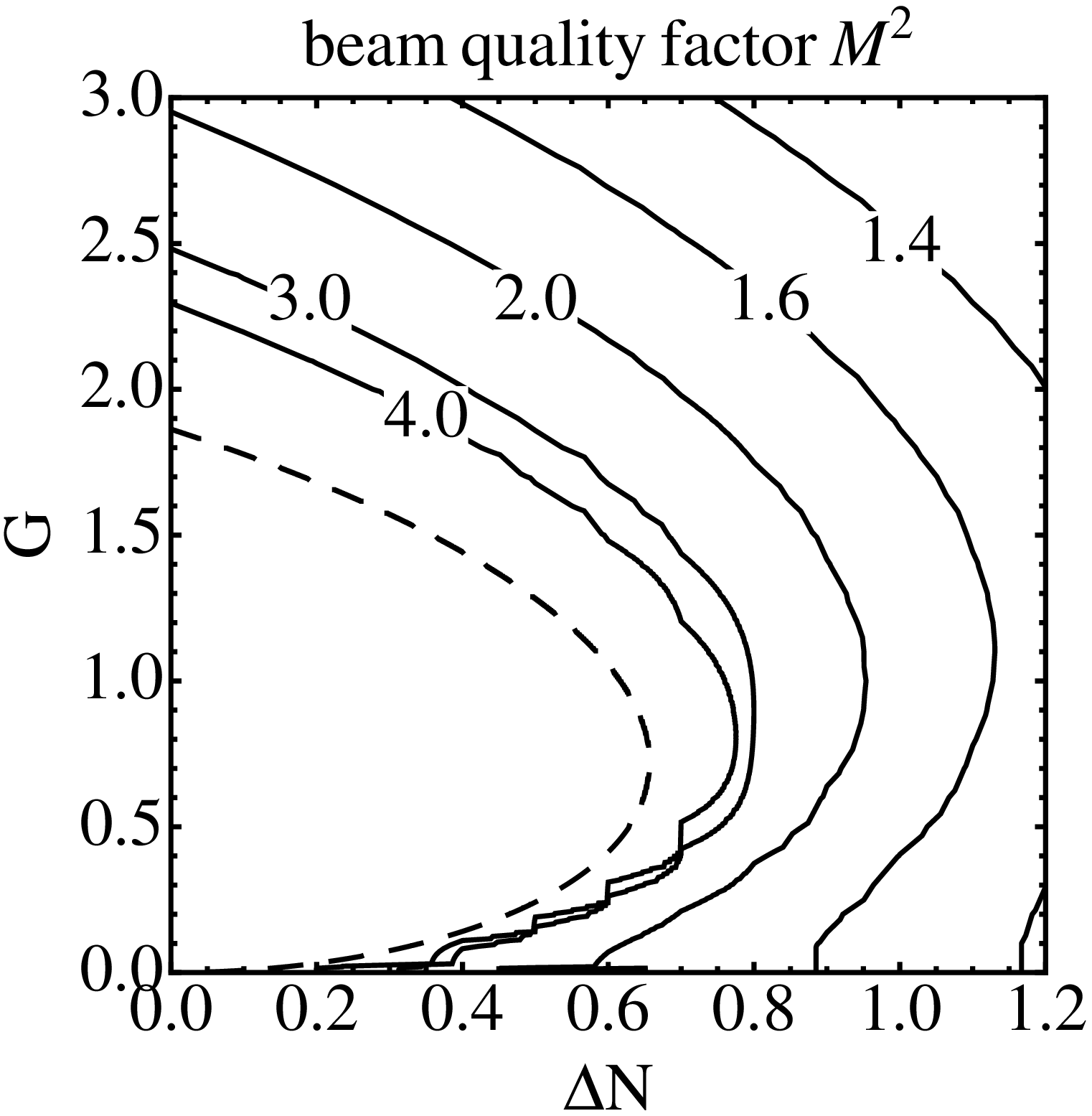}}
    \subfigure[]{\label{fig:M23}\includegraphics[width=2.5in]{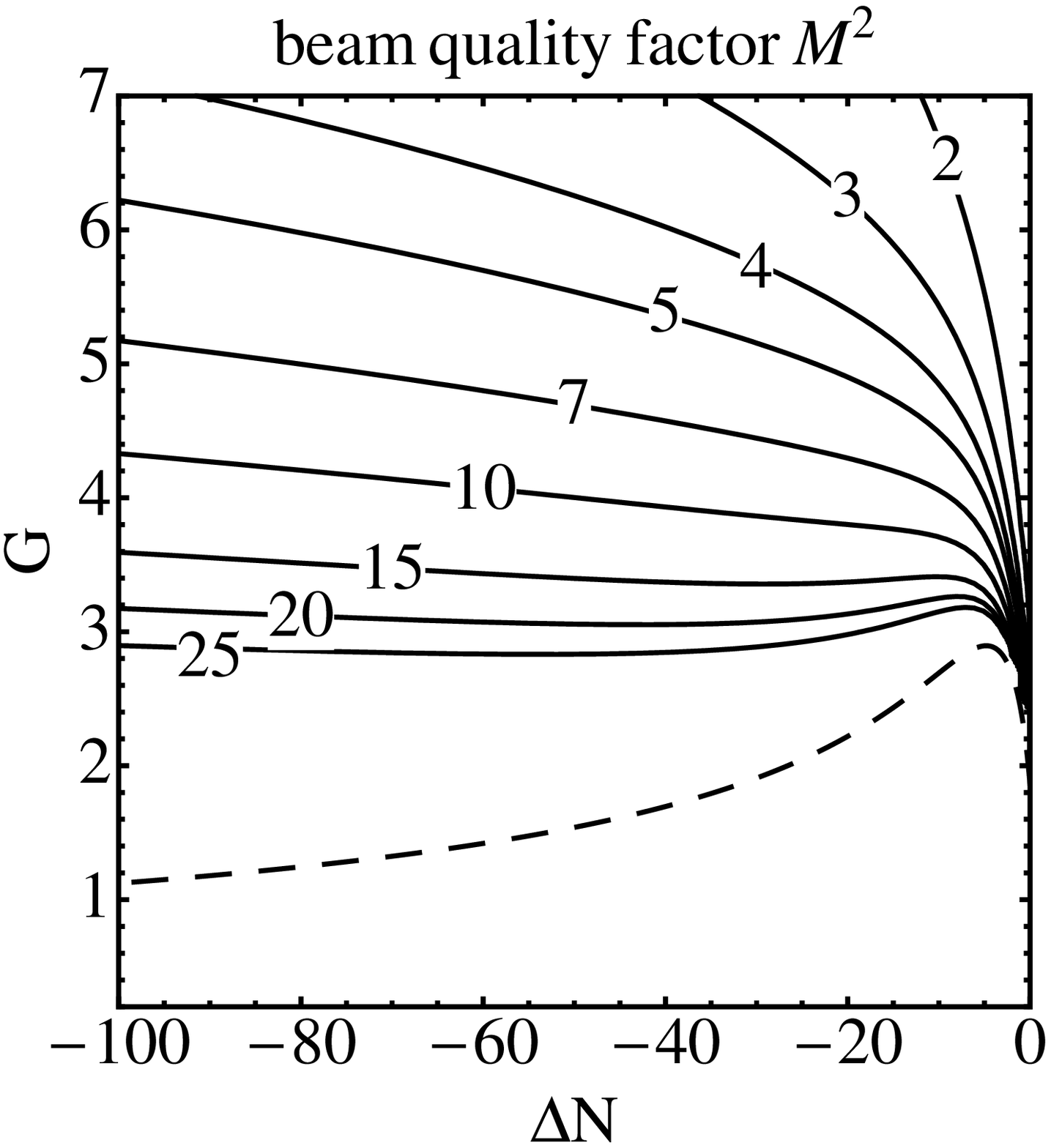}}
    \subfigure[]{\label{fig:M24}\includegraphics[width=2.6in]{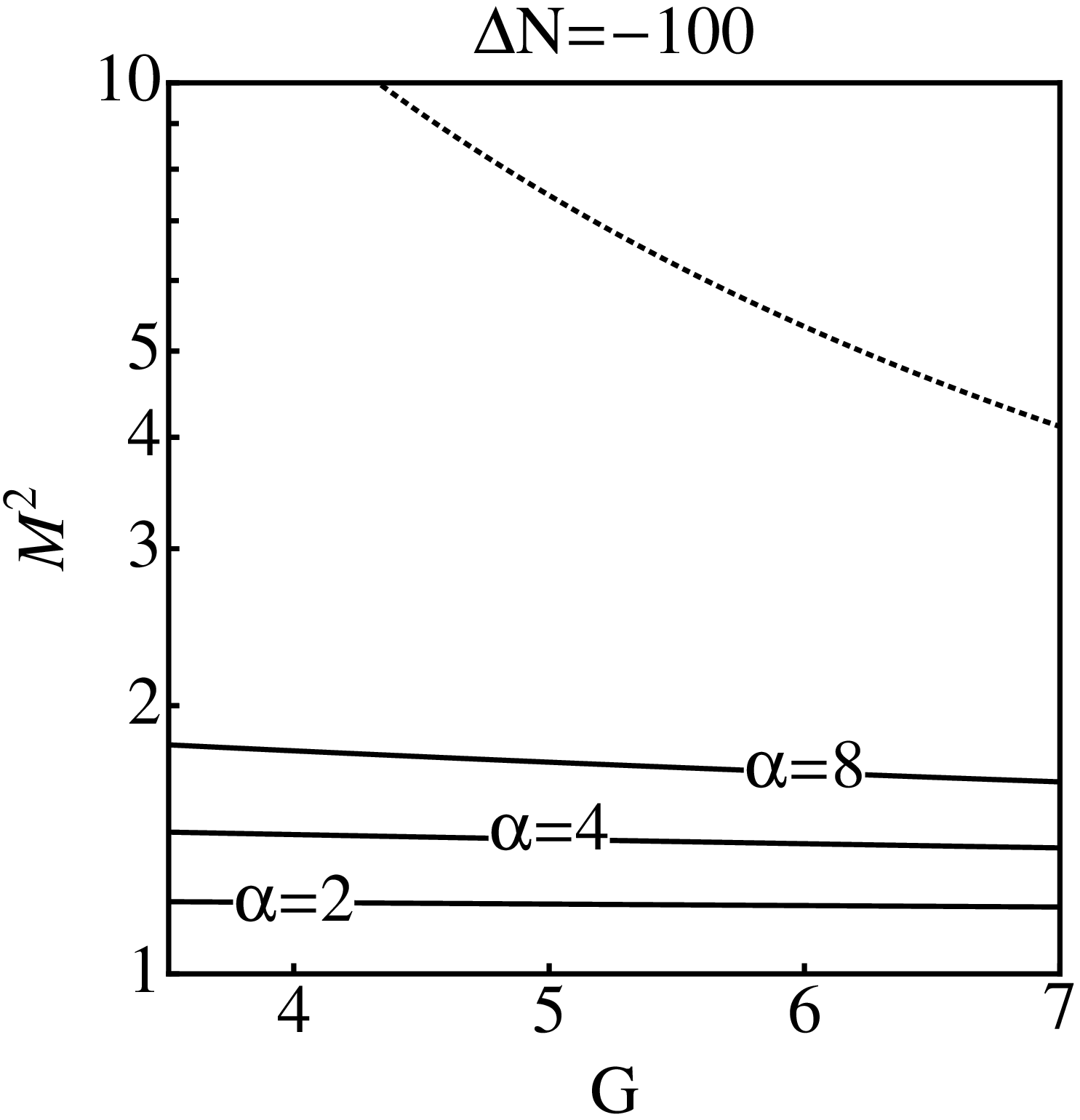}}
  \end{center}
  \caption{Contour plot of $M^2$ as a function of $\Delta N$ and $G$ in the a) GG+IG region,
           b) GG+IG region but zoomed in near the origin, and
           c) GG+IAG region.
            The dashed lines represent the LP01 guiding threshold.  
           d) $M^2$ as a function of $G$ for $\Delta N=-100$. The solid lines relate to the apodized 
              beam and are marked by different values of $\alpha$ from the apodization function.
              The dotted line is in the absence of the apodization function.
              }
\end{figure}

Consider an optical beam with the electric field profile $E(x,y,z)$
propagating in the $z$ direction.
The beam center $\langle x\rangle$ and the 
standard deviation of the intensity distribution $\sigma_x^2$ across the $x$
coordinate are 
\begin{align}
\langle x\rangle (z)&=
\iint dS\ x\ \left|E(x,y,z)\right|^2,\\
\label{eq:stddev}
\sigma_x^2(z)&=\iint
dS\ \Big(x-\langle x\rangle (z)\Big)^2\ \left|E(x,y,z)\right|^2.
\end{align}
Since we only consider cylindrically symmetric optical fibers, the results are 
identical for the $y$ coordinate,
and we take the liberty in dropping the $x$ subscript (e.g. $M^2$)
when convenient.
It can be shown that the standard deviation in Eq.~\ref{eq:stddev}, in the paraxial approximation, obeys a 
universal free-space propagation rule of the form
\begin{equation}
\sigma_x^2(z)=\sigma_x^2(z_0)+A\dfrac{\lambda}{2\pi}(z-z_0)+B\dfrac{\lambda^2}{4\pi^2}(z-z_0)^2.
\end{equation}
$z_0$ is the coordinate of the output facet of the fiber 
which does not necessarily coincide with the position of the beam waist. Ref.~\cite{Yoda1} has shown that
\begin{align}
\label{eq:defineA}
A&=i\iint dS \Big(x-\langle x\rangle(z_0)\Big)
\Big[E\dfrac{\partial E^\star}{\partial x}-c.c.\Big],\\
\label{eq:defineB}
B&=\iint dS \left|\dfrac{\partial E}{\partial x}\right|^2+
\dfrac{1}{4}\Big[\iint dS(E\dfrac{\partial E^\star}{\partial x}
-c.c.)\Big]^2,
\end{align}
where $E\equiv E(x,y,z_0)$ is implied in Eqs.~\ref{eq:defineA} and~\ref{eq:defineB}. 
The position of the beam waist ${\tilde z}_{0x}$ and the beam-quality 
factor $M_x^2$ are given by
\begin{align}
{\tilde z}_{0x}&=z_0-\pi A/(\lambda B)\\
M_x^2&=\sqrt{4B\sigma^2_x(z_0)-A^2}.
\end{align}
We note that we slightly differ with Ref.~\cite{Yoda1} in the sign of the frequency term and also the 
definition of the $A$ term.
While the integrals leading the evaluation of the $M^2$ parameter can be evaluated numerically, 
one is faced with the challenge of reliably truncating the infinite integrals. It is difficult 
to choose a reliable truncation radius, a priori, for a desired error.
Therefore, a closed-form analytical expression is highly desirable. 
Using the parameters defined in Eqs.~\ref{eq:paramDefine1} and~\ref{eq:paramDefine2},  
\begin{align}
\label{eq:paramDefine1}
\zeta^m_{n;u}&=(u^2 + u^{\star 2})^m/(u^2 - u^{\star 2})^n,\\
\label{eq:paramDefine2}
\gamma&=u J_{1}(u)/J_0(u),
\end{align}
${\tilde N}$, $A$, $B$ , and $\sigma_x^2(z_0)$ can be expressed as
\begin{align}
\big(4\pi a^2 {\tilde N}^2\big)^{-1} &=
i\big(\zeta^0_{1;u} + \zeta^0_{1;w}\big).{\rm Im}\big(\gamma\big),
\\
\big(\pi a^2 {\tilde N}^2\big)^{-1}
A&=
-4\big(\zeta^1_{2;u} + \zeta^1_{2;w}\big).{\rm Im}\big(\gamma\big)
-i\big(\zeta^1_{1;u} - \zeta^1_{1;w}\big)
-2i\big(\zeta^0_{1;u} + \zeta^0_{1;w}\big).\left|\gamma\right|^2
\\
\big(2\pi {\tilde N}^2\big)^{-1}
B&=
{\rm Re}\left[\big(u^{\star 2}\zeta^0_{1;u} - w^{\star 2}\zeta^0_{1;w}\big).\gamma\right],
\\
\big(\pi a^4 {\tilde N}^2\big)^{-1}
\sigma_x^2(z_0) &=
-8i\big(\zeta^1_{3;u} - \zeta^1_{3;w}\big).{\rm Im}\big(\gamma\big)
+\big(2\pi a^2 {\tilde N}^2\big)^{-1}
+2\big(\zeta^1_{2;u} + \zeta^1_{2;w}\big)
\\
&\ +4\big(\zeta^0_{2;u} - \zeta^0_{2;w}\big).\left|\gamma\right|^2
\nonumber
\end{align}
The contour plots of $M^2$ as a function of $\Delta N$ and $G$ are shown in 
Figs.~\ref{fig:M21},~\ref{fig:M22},~\ref{fig:M23}. The dashed lines in these figures
identify the threshold values of the dimensionless
gain parameter $G$ required to produce a confined and amplifying
LP01 mode as described in Ref.~\cite{Siegman1}.
The results in Fig.~\ref{fig:M21},
related to the GG+IG region, show that $M^2$ remains very close 
to unity over most of the parameter space. 
A typical single-mode fiber laser operates at $4\lesssim\Delta N\lesssim 6$, for which $M^2\lesssim 1.04$,
regardless of the value of the $G$ parameter. This is consistent with the
common intuition that the beam quality of a single-mode fiber laser is very good and
does not degrade by pumping, unless higher order modes are excited. However, the beam quality factor rapidly increases as
the ${\tilde V}$-parameter of the fiber gets closer to the (dashed) LP01 threshold.
This can be seen in Fig.~\ref{fig:M22} which is similar to Fig.~\ref{fig:M21} but zoomed in near the origin 
and the close proximity of the LP01 threshold line. 

The situation is quite different in the GG+IAG region as shown in Fig.~\ref{fig:M23}.
$M^2$ is quite large over the entire parameter space and 
becomes exceptionally large near the (dashed) LP01 threshold. The large value of the 
$M^2$ especially near the threshold is the result of the long tail of the beam
intensity extending all the way into the cladding region over a large portion of the GG+IAG parameter space~\cite{Siegman2}.
The situation closely resembles the case of a conventional IG fiber ($G=0$), where the tails of the beam 
extend to the cladding in the weakly guiding limit of $\Delta\to 0$, resulting in large values
of $M^2$ as shown in Fig.~\ref{fig:M22}.
We note that the numerical evaluation of $M^2$
becomes increasingly difficult in the regions of large $M^2$ and proper care must be taken 
in the sampling and truncation of the beam to get an accurate result. 

In order to explore the impact of the long intensity tail in the cladding on the value of
$M^2$, we introduce a Gaussian apodization function of the form 
$\exp(-\alpha^{-1}r^2/a^2)$
to softly truncate the long intensity tail of beam. We multiply
the beam profile of Eq.~\ref{eq:E_definition}
by the apodization function
and calculate the $M^2$ numerically. The results are presented in
Fig.~\ref{fig:M24} where $M^2$ is plotted as a function of $G$ for $\Delta N=-100$. The
dotted line is the value of $M^2$ in the absence of the apodization function. 
The three solid lines represent the calculated values of
$M^2$ for the apodized beam and are marked by different values of $\alpha=2, 4, 8,$ from
the apodization function.
Fig.~\ref{fig:M24} clearly shows that the truncation of the intensity 
tail of the GG+IAG beams results in a substantial reduction in the calculated value of $M^2$.

The above observations are quite important in relating the $M^2$ values reported in this paper, especially
for the GG+IAG region of Fig.~\ref{fig:M23} to the experimental measurements. In practice, 
the extended tail of the GG+IAG beam is truncated at the cladding-jacket (or cladding-air)
interface. For example, for $\Delta N=-100$ and $G$ equal to twice the LP01 threshold
value, the field amplitude drops only by 25\% from $r=a$ to $r=2a$; for $\Delta N=-1000$,
this drop in amplitude is only 13\%. Therefore, the impact of the cladding-jacket interface
on the tail of beam warrants close attention. 
\section{Conclusions}
We have shown that $M^2$ can be substantially larger than 
unity even for single-mode fibers in the GG+IAG region.
The large value of $M^2$ in the GG+IAG region
is the result of the long tail of the beam extending all the way 
into the cladding region, the truncation of which 
can lower $M^2$ substantially.
The reported values of $M^2$ in several experiments on GG+IAG fiber lasers
are in the range of 1.05-2.0~\cite{Siegman5,McComb1,Chen1,Chen2}. 
For example, Ref.~\cite{Siegman5} reports $M^2\le 2$ for a GG+IAG optical 
fiber with a 100 $\mu m$ core diameter and a 250 $\mu m$ cladding diameter.
Similarly, Ref.~\cite{McComb1} reports $1.2\le M^2\le 1.5$, where the core and cladding diameters
are 200~$\mu m$ and 340~$\mu m$, respectively.
The cladding diameter is not much larger than the core diameter in
either experiments. Consequently, the long cladding tail of the LP01 beam is truncated at 
the cladding-jacket interface over a large portion of the $\Delta N$-$G$ parameter space 
in the GG+IAG region. Therefore, it is not surprising that these experiments measure such low 
values of $M^2$.

We expect that very large values of $M^2$ will be observed in GG+IAG single-mode fibers
with sufficiently large cladding to core diameter ratios. Measurement of $M^2$ can be challenging
for these fibers since the presence of noise can impact an accurate assessment of the  
tail intensity. The existing experiments~\cite{Siegman5,McComb1,Chen1,Chen2} deviate from the pure GG+IAG formalism of Siegman~\cite{Siegman1} 
in the sense that the cladding-core diameter ratios are not sufficiently large. Therefore, 
an accurate $M^2$ comparison with these experiments is only possible if the 
presence of the cladding-jacket interface is explicitly taken into account. 
Besides the impact on the value of $M^2$,
one must examine whether the truncated mode remains faithful to its GG+IAG nature or is modified into 
a mode which is primarily index-guided by the cladding-jacket index step.

The analytical results of this paper have been independently verified by direct numerical computation.
In order to facilitate the implementation of the analytical results in this paper
for the interested reader, we present the numerical values of the 
analytical solution at ${\tilde V}^2=-5+7i$. We get: 
$w=0.79722+3.04536i$, ${\tilde N}a=0.27742$,
$A=-2.01296$, $Ba^2=2.5852$, $\sigma_x^2(z_0)=0.58852a^2$, and $M^2=1.42611$.

\end{document}